\documentclass[prb,twocolumn,aps,showpacs,superscriptaddress,epsfig]{revtex4-2}

\usepackage{amssymb}
\usepackage{amsbsy}
\usepackage{amsmath}
\usepackage{amsfonts}
\usepackage{bm}
\usepackage{physics}
\usepackage[dvipsnames,table]{xcolor}
\usepackage{epsfig}
\usepackage[bb=boondox]{mathalfa}
\usepackage{graphicx}
\usepackage{subfigure}
\usepackage{dsfont}
\usepackage{array}
\usepackage{mathrsfs}
\usepackage{float}
\usepackage{hyperref}
\usepackage[toc,page]{appendix}

\newcommand{\ang}{\AA}

\hypersetup{
    colorlinks=true,
    linkcolor=blue,
    filecolor=cyan,
    urlcolor=magenta,
    citecolor=red,
}

\newcommand{\supplement}[1]{%
  \clearpage%
  \onecolumngrid%
  \title{#1}%
  \maketitle%
  \setcounter{section}{0}%
  \setcounter{equation}{0}%
  \setcounter{figure}{0}%
  \setcounter{table}{0}%
  \setcounter{page}{1}%
  \renewcommand{\thesection}{S\arabic{section}}%
  \renewcommand{\thesubsection}{\Alph{subsection}}%
  \renewcommand{\theequation}{S\arabic{equation}}%
  \renewcommand{\thefigure}{S\arabic{figure}}%
  \renewcommand{\thetable}{S\Roman{table}}%
  \renewcommand{\thepage}{S\arabic{page}}%
  \numberwithin{figure}{section}%
  \numberwithin{table}{section}%
  \numberwithin{equation}{section}%
}

\makeatletter
\def\maketitle{
\@author@finish
\title@column\titleblock@produce
\suppressfloats[t]}
\makeatother
\makeatletter
\newcommand{\suppTOCon}{%
  \let\old@addcontentsline\addcontentsline
  \renewcommand{\addcontentsline}[3]{%
    \edef\@tempa{##1}\edef\@tempb{toc}%
    \ifx\@tempa\@tempb
      \old@addcontentsline{stoc}{##2}{##3}%
    \else
      \old@addcontentsline{##1}{##2}{##3}%
    \fi
  }%
}
\newcommand{\suppTOCoff}{%
  \let\addcontentsline\old@addcontentsline
}
\makeatother

\begin{document}

\title{Nonlinear Hall Effect in Metal--Organic Frameworks}

\author{Sarbajit~Mazumdar}
\email{sarbajit.mazumdar@uni-wuerzburg.de}
\affiliation{Institut f\"ur Theoretische Physik und Astrophysik and W\"urzburg-Dresden Cluster of Excellence ctd.qmat,
Universit\"at W\"urzburg, Am Hubland Campus S\"ud, 97074 W\"urzburg, Germany}

\author{Jagadish~N~S}
\affiliation{Centre for Condensed Matter Theory, Department of Physics, Indian Institute of Science, Bengaluru, India - 560012}

\author{Awadhesh~Narayan}
\affiliation{Solid State and Structural Chemistry Unit, Indian Institute of Science, Bangalore 560012, India}

\author{Giorgio~Sangiovanni}
\affiliation{Institut f\"ur Theoretische Physik und Astrophysik and W\"urzburg-Dresden Cluster of Excellence ctd.qmat,
Universit\"at W\"urzburg, Am Hubland Campus S\"ud, 97074 W\"urzburg, Germany}

\author{Ronny~Thomale}
\affiliation{Institut f\"ur Theoretische Physik und Astrophysik and W\"urzburg-Dresden Cluster of Excellence ctd.qmat,
Universit\"at W\"urzburg, Am Hubland Campus S\"ud, 97074 W\"urzburg, Germany}

\author{Arka~Bandyopadhyay}
\email{arka.bandyopadhyay@uni-wuerzburg.de}
\affiliation{Institut f\"ur Theoretische Physik und Astrophysik and W\"urzburg-Dresden Cluster of Excellence ctd.qmat,
Universit\"at W\"urzburg, Am Hubland Campus S\"ud, 97074 W\"urzburg, Germany}

\date{\today}

\begin{abstract}
We propose metal--organic frameworks (MOFs) as tunable platforms for nonlinear Hall responses. A universal analytical downfolding scheme maps $C_3$-symmetric frameworks onto star- and honeycomb-lattice models, reproducing first-principles Dirac features. Spin--orbit coupling and broken inversion symmetry gap the Dirac cones, generating Berry-curvature hot spots. Symmetry analysis identifies tailored synthetic pathways, including linker design, as intrinsic routes to engineer nonlinear Hall transport beyond strain and substrate control.
\end{abstract}

\maketitle

\textit{Introduction.--}
The nonlinear Hall effect (NLHE) is an unconventional transverse response that can arise in time-reversal-invariant crystals when inversion symmetry is broken. Unlike the conventional anomalous Hall effect, where the Berry curvature of Bloch bands governs the linear Hall response as an effective magnetic field in momentum space~\cite{PhysRevLett.45.494,RevModPhys.82.1959,RevModPhys.82.1539}, the NLHE originates from an asymmetric Berry-curvature distribution across the Brillouin zone~\cite{PhysRevLett.115.216806, du2021nonlinear,ortix2021nonlinear,bandyopadhyay2024non,ma2019observation, kang2019nonlinear,makushko2024tunable,min2023strong,sinha2022berry,ye2023control,Nomoto2025}. This asymmetry, encoded in the Berry-curvature dipole, enables a second-order Hall response even in time-reversal-symmetric systems. Its relevance for rectification~\cite{kumar2021room,wang2019ferroicity}, photodetection~\cite{kim2019prediction}, and photovoltaic responses~\cite{PhysRevLett.119.067402} motivates the search for platforms where symmetry and band geometry can be engineered.

\begin{figure*}
    \centering
    \includegraphics[width=1\textwidth]{Figures/final_new_figure_1.jpg}
    \vspace{-1\baselineskip}
\caption{\textbf{From Cu--DCA to the effective star lattice.}
(a) Relaxed Cu--DCA monolayer. Inset: the three N atoms around each Cu site form an effective triangle after Cu decimation.
(b) Orbital-projected bands showing dominant C/N $p_z$ character near the Fermi level.
(c) Linker decimation: eliminated sites are marked in red, yielding renormalized terminal hoppings between $L$ and $R$.
(d) Effective six-site star lattice spanned by $\vec a_1$ and $\vec a_2$, with intra-triangle hoppings $t_{\triangle}$ and $t_{\bigtriangledown}$, inter-triangle hopping $t$ modified to $t_\perp$ under uniaxial anisotropy, and intrinsic Kane--Mele SOC $\pm i\lambda_{\rm SO}$ shown by green arrows.}
    \label{fig:1}
\end{figure*}

Metal--organic frameworks (MOFs), composed of metal nodes linked by organic ligands~\cite{zhou2012introduction,furukawa2013chemistry}, offer such control through their structural and chemical tunability. Their electronic bands can be engineered by choosing metal centers and organic linkers, making MOFs natural platforms for designer lattice models~\cite{jiang2021exotic}. In particular, two-dimensional (2D) MOFs have been predicted to host Dirac and flat bands, as well as topological-insulator and Chern-insulator phases~\cite{wang2013organic,PhysRevLett.110.196801,wang2013prediction,Hu2023_NatCommun,tin2023haldane}. Nevertheless, their potential for NLHE remains largely unexplored.

In this Letter, we introduce a downfolding framework that maps MOFs onto effective lattice models for designing strong and controllable NLHE. Focusing on $C_3$-symmetric 2D MOFs, we show that the experimentally relevant Cu-dicyano-anthracene (Cu--DCA) monolayer~\cite{fuchs2020kagome, Zhang2016_NanoLett_DCA_TI,Kumar2018_NanoLett_MOF_Bands}, pyrazine–metal and triphenyl-metal (metal = Pb/Bi) MOF~\cite{wang2013organic,wang2013quantum} can be mapped onto effective star and honeycomb lattice descriptions through real-space decimation~\cite{bandyopadhyay20218,bandyopadhyay2020review,banerjee2024non,bose2025origin}. The resulting hopping parameters reproduce the distinct first-principles band topologies, including gapped Dirac cones that provide a natural source of Berry-curvature hot spots. This mapping provides a route to connect chemical and structural tunability with symmetry-selective band geometry, identifying strain engineering and symmetry-lowered sister compounds as practical pathways to finite Berry curvature, a tunable Berry-curvature dipole, and sizable NLHE in MOFs.

\textit{Downfolding of Cu--DCA.--}
We begin with the experimentally synthesized Cu--DCA MOF, shown in Fig.~\ref{fig:1}(a)~\cite{kumar2021manifestation,zhang2014probing,hernandez2021searching,yan2021two,yan2021synthesis}. Its first-principles band structure [Fig.~\ref{fig:1}(b)] is semimetallic and hosts Dirac crossings at the high-symmetry $K$ point~\cite{zhang2016intrinsic,fuchs2020kagome}. Since gap opening at such crossings produces Berry-curvature hot spots, Cu--DCA is a natural starting point for NLHE engineering once SOC is included and inversion symmetry is broken.

The orbital-resolved bands show that the low-energy states are dominated by C/N $p_z$ orbitals, while Cu contributes weakly near the Fermi level [Fig.~\ref{fig:1}(b)]. This motivates a single-orbital C/N description. We first decimate the Cu-centered coordination environment into effective hoppings among the three surrounding N atoms, forming equilateral triangles [inset of Fig.~\ref{fig:1}(a)]. A second decimation integrates out the linker segment between neighboring triangles, yielding the star-lattice geometry with renormalized hoppings [Fig.~\ref{fig:1}(c)]. The analytical flow and parameters are summarized in Supplementary Sec.~\ref{sec:method_decimation}. Thus, Cu--DCA is reduced to a compact effective model retaining the Dirac crossings and low-energy band geometry relevant for the NLHE.

\textit{Dipolar component of Berry curvature.--}
The downfolded lattice in Fig.~\ref{fig:1}(d) has a six-site unit cell formed by two inequivalent triangular motifs, with intra-triangle hoppings $t_{\triangle}$ and $t_{\bigtriangledown}$. Neighboring triangles are connected by $t$ in the pristine lattice and by $(t,t_\perp)$ under uniaxial anisotropy. The effective Hamiltonian is
\begin{equation}
\mathcal{H}=\mathcal{H}_0+\mathcal{H}_{\delta}+\mathcal{H}_{\rm SO},
\label{eq:ham_total}
\end{equation}
where $\mathcal{H}_0$ contains intra- and inter-triangle hoppings, $\mathcal{H}_{\delta}$ encodes inversion-breaking onsite offsets between triangular subunits, and $\mathcal{H}_{\rm SO}$ is a Kane--Mele-type intrinsic SOC term. The full real-space form is given in Supplementary Sec.~\ref{sec:supp_hamiltonian}.

Representative band structures of the star lattice without SOC are shown in Supplementary Fig.~\ref{fig:band_struc}. We focus on the parameter regime in which two kagome-like band manifolds are energetically separated, since this is the regime that reproduces the first-principles electronic structure of Cu--DCA. Once SOC and inversion-symmetry breaking are included, gaps open at the Dirac crossings. These gap openings are the source of sharply localized Berry-curvature hot spots and, once the remaining point-group symmetry is broken appropriately, of a finite dipolar component.

To make that connection explicit, we first analyze the Berry-curvature distribution of the effective model. For a two-dimensional Bloch band $n$, the only nonvanishing component is $\Omega_n^z(\mathbf{k})$, which can be written as
\begin{equation}
\Omega^z_{n}(\mathbf{k})=
-2\,\mathrm{Im}
\sum_{n'\neq n}
\frac{
\langle u_{n\mathbf{k}}|v_x|u_{n'\mathbf{k}}\rangle
\langle u_{n'\mathbf{k}}|v_y|u_{n\mathbf{k}}\rangle
}{
\left(E_n(\mathbf{k})-E_{n'}(\mathbf{k})\right)^2
}.
\label{eq:berry}
\end{equation}

\begin{figure}
    \centering
    \includegraphics[width=1\linewidth]{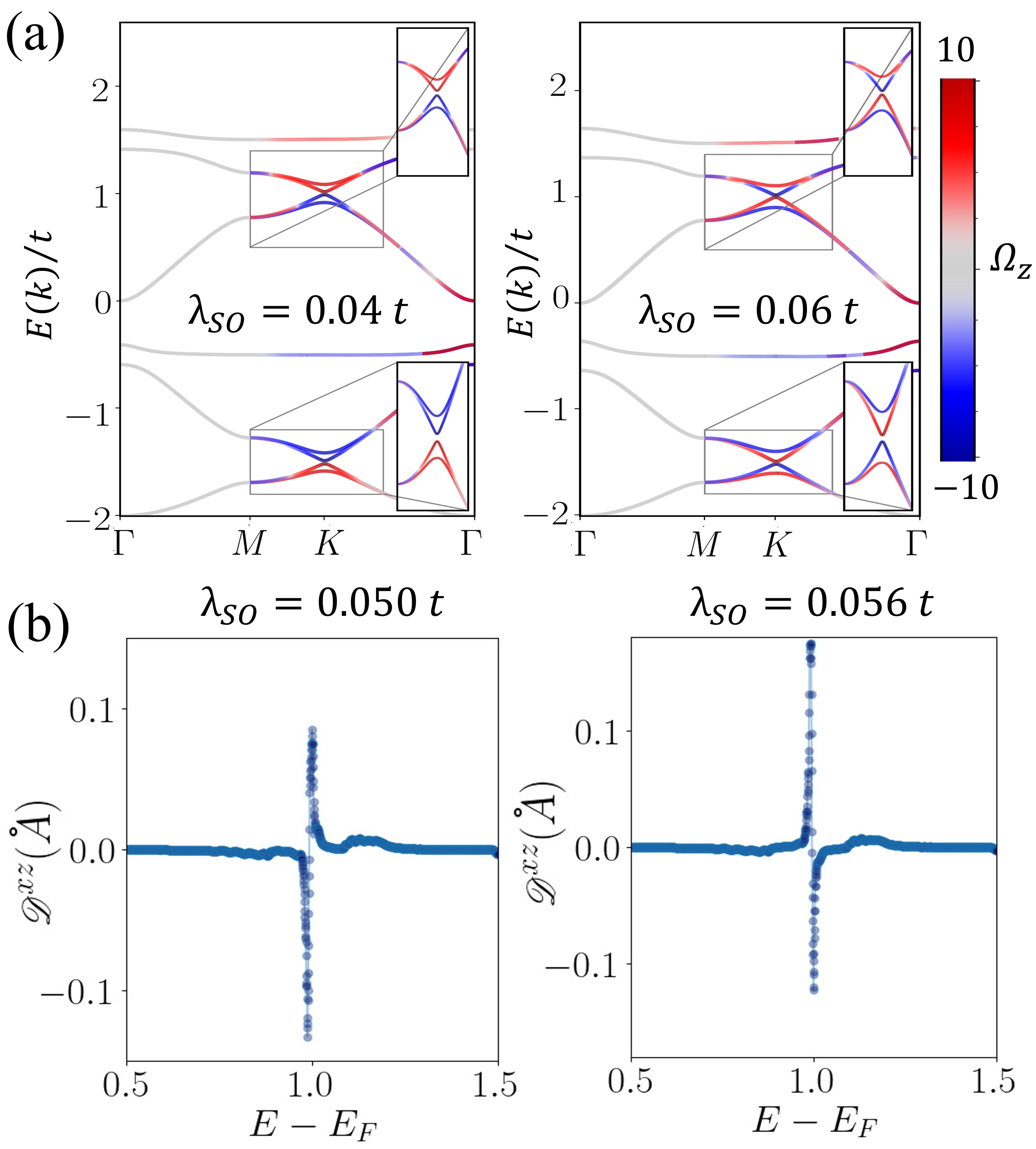}

\caption{\textbf{Berry curvature and Berry-curvature dipole in the star lattice.}
(a) Berry-curvature-projected bands of the onsite-modulated model with $\delta_1=-0.08\,t$, $\delta_2=0.08\,t$, and $t_{\triangle}=t_{\bigtriangledown}=0.5\,t$ for two SOC strengths. Increasing $\lambda_{\rm SO}$ drives a band inversion near $K$ and reverses the Berry-curvature hot spots.
(b) Corresponding $D^{xz}$ versus Fermi level with small anisotropy $t_\perp\neq t$, modeling $2\%$ tensile strain. The strong enhancement and sign reversal near the critical point reflect the topological transition.}
    \label{fig:5}
\end{figure}

Time-reversal symmetry imposes $\Omega_n^z(\mathbf{k})=-\Omega_n^z(-\mathbf{k})$, so the linear anomalous Hall response vanishes after Brillouin-zone integration, while the nonlinear response can survive through the first moment of Berry curvature once inversion symmetry is broken. Figure~\ref{fig:5}(a) shows the Berry-curvature-projected bands of the onsite-modulated star lattice with $\delta_1=-0.08\,t$, $\delta_2=0.08\,t$, and $t_{\triangle}=t_{\bigtriangledown}=0.5\,t$. Increasing $\lambda_{\rm SO}$ from $0.04\,t$ to $0.06\,t$ drives a band inversion near $K$ and reverses the Berry-curvature hot spots at the avoided crossings. The associated topological change is confirmed by the $\mathds{Z}_2$ invariant from the Wannier charge center flow; see Supplementary Sec.~\ref{sec:wann} and Fig.~\ref{fig:wcc}. The same mechanism also operates in the breathing version of the star lattice, as shown in Supplementary Sec.~\ref{sec:bc_bcd_supp}.

\begin{figure*}
    \centering
    \includegraphics[width=1\linewidth]{Figures/new_fig.jpg}

\caption{\textbf{Dipolar component of Berry curvature in MOFs.}
(a) DFT band structure of Cu--DCA with two separated kagome-like manifolds; the Fermi level lies near the Dirac point of the upper manifold. Inset: replacing one Cu atom by Zn breaks inversion symmetry.
(b) Dipolar-component phase diagram obtained from DFT-consistent hoppings ($t_{\triangle}=t_{\bigtriangledown}=-0.11$~eV, $t=0.81$~eV, $t_\perp=-0.78$~eV) in the presence of SOC, inversion-symmetry breaking, and $2\%$ uniaxial strain. The dipolar component changes sign across the topological transition.
(c) Synthesis-based route to intrinsic symmetry lowering by modifying one of the three equivalent linker directions.
(d) Generalization to a pyrazine--metal MOF (Pb shown), which can also be downfolded to a star lattice; because the two Pb sites lie in different planes, a transverse electric field can break inversion symmetry.
(e) Orbital-projected DFT band structure of pyrazine--Pb, showing that the star-lattice-like bands are dominated by C and N states, with no dominant Pb contribution near the Fermi level.}
    \label{fig:figure_DFT}
\end{figure*}
The dipolar component tensor is defined as
\begin{equation}
\mathscr{D}^{ab}
=
\sum_n
\int_{\mathrm{BZ}}
\frac{d^2k}{(2\pi)^2}\,
f^{(n)}_0(\mathbf{k})\,
\partial^a \Omega_n^b(\mathbf{k}),
\label{eq:bcd}
\end{equation}
where $f^{(n)}_0$ is the Fermi--Dirac distribution. Within semiclassical Boltzmann theory, the static nonlinear Hall current is~\cite{PhysRevB.98.121109,PhysRevLett.123.196403,PhysRevB.92.235447,PhysRevLett.115.216806,PhysRevB.107.205124}
\begin{equation}
\vec{\mathcal{J}}_H
=
-\frac{\tau}{2}\,
\hat z \times \boldsymbol{\mathcal{E}}\,
(\mathbf{D}\cdot\boldsymbol{\mathcal{E}}),
\label{eq:nlh_current}
\end{equation}
with $\tau$ the relaxation time and $\mathbf{D}$ the dipolar component vector in two dimensions. When a single mirror plane $\mathcal{M}_x$ is preserved, symmetry enforces $\mathscr{D}^{yz}=0$ while allowing $\mathscr{D}^{xz}\neq0$~\cite{bandyopadhyay2022electrically}. For an in-plane electric field $\boldsymbol{\mathcal{E}}=\mathcal{E}(\cos\theta,\sin\theta)$, the nonlinear Hall response therefore obeys $\mathcal{J}_H=\sigma_H(\theta)|\boldsymbol{\mathcal{E}}|^2$ with $\sigma_H(\theta)=-(\tau/2)\,\mathscr{D}^{xz}\sin\theta$.

The energy-resolved response in Fig.~\ref{fig:5}(b) shows that the dominant contribution to $\mathscr{D}^{xz}$ comes from the SOC-split Dirac crossings identified in Fig.~\ref{fig:5}(a). In the onsite-modulated model, uniaxial $2\%$ strain together with SOC yields a peak value $D^{xz}\sim 0.2~\ang$ near $\lambda_{\rm SO}=0.056\,t$. In the breathing case, the corresponding peak is $D^{xz}\sim 0.05~\ang$ near $\lambda_{\rm SO}=0.04\,t$ [Supplementary Fig.~\ref{fig:3}]. Taking $\tau=10~\mathrm{ps}$, these values translate into nonlinear Hall conductivities of order $\sigma_H\sim 0.037~\mathrm{nm\,S\,V^{-1}}$ and $\sigma_H\sim 0.01~\mathrm{nm\,S\,V^{-1}}$, respectively; see Supplementary Sec.~\ref{nlh}. The sharp enhancement and sign reversal of the dipolar component across the transition are thus direct consequences of SOC-driven band inversion in the effective star lattice.

\textit{Material implementation and design principles.--}
Having identified the band-geometry mechanism in the effective model, we now return to the material level. Figure~\ref{fig:figure_DFT}(a) shows that the DFT band structure of Cu--DCA indeed contains two energetically separated kagome-like band manifolds, with the Fermi level lying close to the Dirac point of the upper one. Using the DFT-consistent hoppings $t_{\triangle}=t_{\bigtriangledown}=-0.11$~eV, $t=0.81$~eV, and $t_\perp=-0.78$~eV, we compute the dipolar component phase diagram shown in Fig.~\ref{fig:figure_DFT}(b). The response is maximized near the topological phase boundary, where the SOC-induced avoided crossing produces strongly concentrated Berry curvature and a sign change of the dipolar component. This demonstrates that the effective star-lattice description captures not only the band structure but also the symmetry-controlled nonlinear response of the MOF.

In Cu--DCA, inversion symmetry can be broken by an appropriate substrate, by heterostructure engineering, or by chemical substitution, as illustrated by the Zn-doped configuration in the inset of Fig.~\ref{fig:figure_DFT}(a). In addition, the uniaxial anisotropy required to expose a nonzero $\mathscr{D}^{xz}$ need not be supplied externally. Figure~\ref{fig:figure_DFT}(c) sketches a synthesis-based route in which one of the three equivalent linker directions is modified, for example by directly connecting two anthracene-like units or by inserting an acetylenic bridge~\cite{cirera2020tailoring,kawai2018diacetylene}. In the decimated description, this changes the effective hopping between the terminal sites $L$ and $R$ of Fig.~\ref{fig:1}(c) from $t=-{\lambda'}^{2}/\lambda$ to an anisotropic value $t_\perp=(t_3/\lambda)t$ along the selected direction, where $t_3$ is the hopping amplitude through the modified bridge. The resulting structure lowers the original $C_3$ symmetry to a single mirror plane and simultaneously removes inversion, thereby generating the finite Berry curvature and the symmetry-allowed dipolar component needed for the NLHE without externally applied strain.

Using Eq.~\eqref{eq:nlh_current}, we estimate the experimentally observable response for two symmetry-lowered realizations: onsite modulation and breathing-type intra-triangle hopping modulation. For an in-plane electric field of magnitude $\mathcal{E}\sim10^4~\mathrm{V/m}$, the expected nonlinear Hall current reaches $\mathcal{J}_H\sim 0.037~\mathrm{mA/cm}$ in the onsite-modulated case and $\mathcal{J}_H\sim 0.01~\mathrm{mA/cm}$ in the breathing case (Supplementary Sec.~\ref{nlh}). These values fall within current experimental sensitivity.

The design principle is not restricted to Cu--DCA. As illustrated in
Fig.~\ref{fig:figure_DFT}(d), a second class of $C_3$-symmetric
MOFs, exemplified by a pyrazine--metal (Pb) MOF, which may be viewed
as a nitrogenated triphenyl--Pb framework [Fig.~\ref{fig:s5}(a)], can
likewise be downfolded to an effective star-lattice model, with an
intertriangle hopping amplitude $t^{\prime}=-2t$. In this structure, the two Pb sites in the unit cell reside in different planes, giving
rise to an intrinsically buckled geometry. A transverse electric field, for example, induced by an underlying substrate, can then lift the
remaining inversion symmetry in a controlled manner. When combined with
SOC, this controlled inversion-symmetry breaking provides the essential
ingredients for generating finite Berry curvature and the associated
Berry-curvature-dipole response.

The pristine DFT band structure shown in Fig.~\ref{fig:figure_DFT}(e)
closely follows the characteristic star-lattice dispersion, thereby
validating the downfolded description. By contrast, the
orbital-projected band structure of the triphenyl--Pb framework
[Fig.~\ref{fig:s5}(b)] shows that all constituent atoms contribute
significantly to the Dirac-like bands near the Fermi level, indicating
that its appropriate low-energy effective description is a honeycomb
lattice rather than a star lattice. Nevertheless, in both cases, the
same symmetry mechanism identified for Cu--DCA--namely, the selective
extension of one-third of the nearest-neighbor intertriangle linkages
through suitable linkers, should generate a finite NLHE in this broader
class of MOFs.

More generally, any $C_3$-symmetric MOF in which a metal center
connects three symmetry-related nonmetal sites provides a natural
starting point for realizing an effective star-lattice description and,
consequently, tunable NLHE physics. Related MOF platforms, such as
Ni$_2$C$_{24}$S$_6$H$_{12}$~\cite{wei2016spin}, as well as suitably
designed covalent organic frameworks~\cite{yan2026engineering,yang2026breathing}
and conjugated polymer networks~\cite{adjizian2014dirac}, appear
promising in the same spirit.


\textit{Conclusions.--}
We have shown that two-dimensional MOFs provide a tunable platform for nonlinear Hall physics by combining analytical downfolding, symmetry analysis, and first-principles calculations. For Cu--DCA, real-space decimation yields an effective star lattice that captures the separated kagome-like bands and the Dirac feature near the Fermi level. Inversion-symmetry breaking and SOC then generate Berry-curvature hot spots and a Berry-curvature dipole whose magnitude and sign are controlled by the topological transition. The response can be engineered through substrates, doping, strain, or linker-level synthesis that embeds the required anisotropy. This strategy is universal and extends to broader $C_3$-symmetric MOFs, including pyrazine– and triphenyl–metal networks, establishing MOFs as designable platforms for nonlinear Hall transport.

\textit{Acknowledgments.--}
The authors gratefully acknowledge the Gauss Centre for Supercomputing e.V. (www.gauss-centre.eu) for funding this project by providing computing time on the GCS Supercomputer SuperMUC-NG at Leibniz Supercomputing Centre (www.lrz.de). The authors also acknowledge financial support by the Deutsche Forschungsgemeinschaft through Project--ID 258499086 -- SFB 1170, through the W\"urzburg--Dresden Cluster of Excellence on Complexity and Topology in Quantum Matter -- ctd.qmat, Project-ID 390858490 -- EXC 2147, and through the Research Unit QUAST, Project-ID 449872909 -- FOR5249. A.N. acknowledges support from DST CRG grant (CRG/2023/000114).

\let\oldaddcontentsline\addcontentsline
\renewcommand{\addcontentsline}[3]{}
\bibliography{references}
\let\addcontentsline\oldaddcontentsline

\supplement{Supplemental Material:\\
Nonlinear Hall Effect in Metal--Organic Frameworks}

\tableofcontents

\twocolumngrid

\section{Effective star-lattice model}
\label{sec:supp_model}

\subsection{Hamiltonian}
\label{sec:supp_hamiltonian}

The effective star-lattice Hamiltonian is written as
\begin{equation}
\mathcal{H}=\mathcal{H}_0+\mathcal{H}_{\delta}+\mathcal{H}_{\rm SO}.
\label{eq:supp_ham}
\end{equation}
The spin-independent hopping part reads
\begin{multline}
\mathcal{H}_0=
-t_{\triangle}\sum_{\substack{\langle ij\rangle\in\triangle\\\sigma}}
c_{i\sigma}^{\dagger}c_{j\sigma}
-t_{\bigtriangledown}\sum_{\substack{\langle ij\rangle\in\bigtriangledown\\\sigma}}
c_{i\sigma}^{\dagger}c_{j\sigma} \\
-t\sum_{\substack{\langle ij\rangle\in\triangle\rightarrow\bigtriangledown\\\sigma}}
c_{i\sigma}^{\dagger}c_{j\sigma}
+\mathrm{h.c.},
\label{eq:supp_h0}
\end{multline}
where $t_{\triangle}$ and $t_{\bigtriangledown}$ are the intra-triangle hoppings on the two inequivalent triangular motifs of the star lattice, while $t$ denotes the inter-triangle hopping in the pristine system. A breathing distortion corresponds to $t_{\triangle}\neq t_{\bigtriangledown}$, and uniaxial anisotropy is modeled by replacing one of the inter-triangle hoppings by $t_\perp$.
\begin{figure}
    \centering
    \includegraphics[width=1\linewidth]{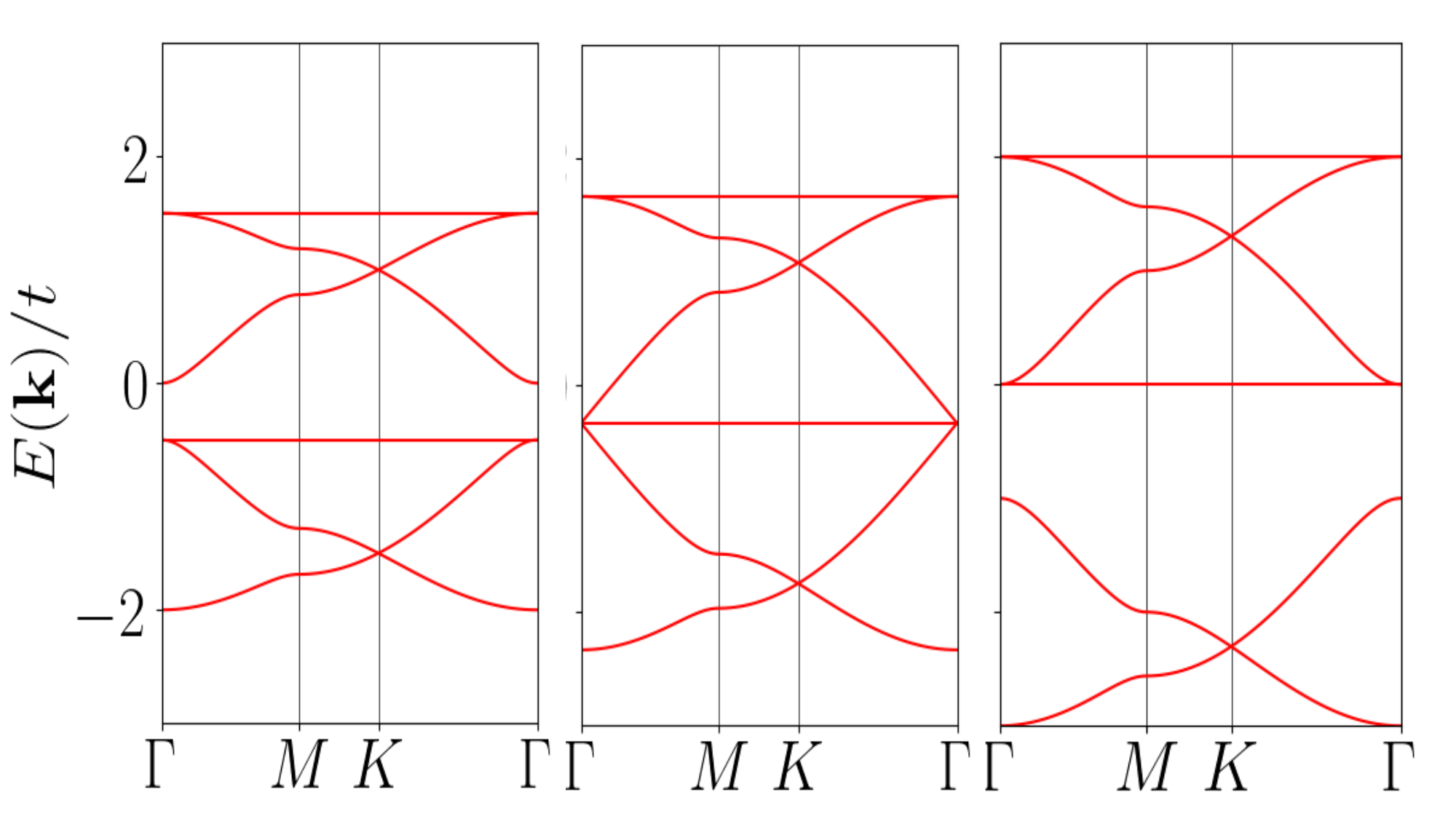}
    \caption{\textbf{Representative band structures of the pristine star lattice.}
    Band structures along $\Gamma\rightarrow M\rightarrow K\rightarrow\Gamma$ for $t_{\triangle}=t_{\bigtriangledown}=0.5\,t$, $t_{\triangle}=t_{\bigtriangledown}=\frac{2}{3}t$, and $t_{\triangle}=t_{\bigtriangledown}=t$. Throughout the calculations, we set $t=1$~eV.}
    \label{fig:band_struc}
\end{figure}
Inversion-symmetry breaking through onsite modulation is captured by
\begin{equation}
\mathcal{H}_{\delta}
=
\delta_1\sum_{\substack{i\in\triangle\\\sigma}} c_{i\sigma}^{\dagger}c_{i\sigma}
+
\delta_2\sum_{\substack{i\in\bigtriangledown\\\sigma}} c_{i\sigma}^{\dagger}c_{i\sigma},
\label{eq:supp_delta}
\end{equation}
where $\delta_1$ and $\delta_2$ are onsite energies on the two triangular subunits.

The intrinsic spin--orbit coupling is taken in Kane--Mele form,
\begin{equation}
\mathcal{H}_{\rm SO}
=
i\lambda_{\rm SO}\sum_{\langle\!\langle ij\rangle\!\rangle,\alpha,\beta}
\nu_{ij}\,
c_{i\alpha}^{\dagger}
(s^z)_{\alpha\beta}
c_{j\beta}
+
\mathrm{h.c.},
\label{eq:supp_soc}
\end{equation}
where $\lambda_{\rm SO}$ is the SOC strength, $\nu_{ij}=\pm1$ encodes the chirality of the hopping path, and $(s^z)_{\alpha\beta}$ is the spin matrix.

\subsection{Representative band structures}
\label{sec:supp_band_evolution}

Figure~\ref{fig:band_struc} shows representative band structures of the pristine star lattice along the high-symmetry path $\Gamma\rightarrow M\rightarrow K\rightarrow\Gamma$. By tuning the ratio between intra-triangle and inter-triangle hoppings, the model interpolates between dispersions with different degrees of kagome-like character. The regime emphasized in the main text is the one with two well-separated kagome-like band manifolds, because this reproduces the DFT band structure of Cu--DCA most closely and places the Fermi level near Dirac crossings that can be converted into Berry-curvature hot spots once SOC and inversion-symmetry breaking are introduced.
\begin{figure}
    \centering
    \includegraphics[width=1\linewidth]{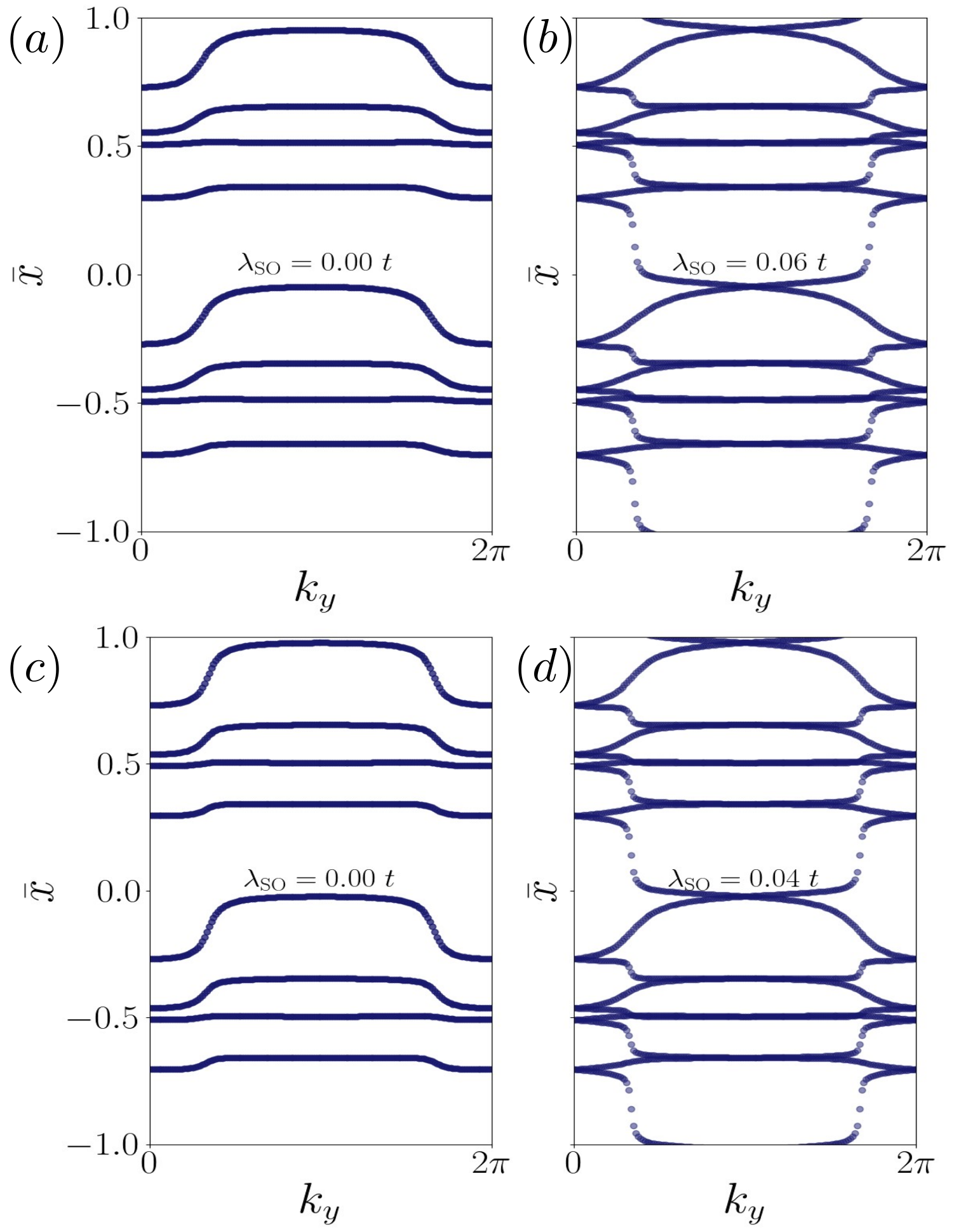}
    \caption{\textbf{Wannier charge centers and topological phase transition.}
    Upper panels (a,b): WCC evolution for the onsite-modulated model with $\delta_1=-0.08\,t$, $\delta_2=0.08\,t$, and $t_{\triangle}=t_{\bigtriangledown}=0.5\,t$.
    Lower panels (c,d): WCC evolution for the breathing star lattice with $t_{\triangle}=0.45\,t$ and $t_{\bigtriangledown}=0.5\,t$.
    In both cases, increasing $\lambda_{\rm SO}$ drives a $\mathds{Z}_2$ transition. An odd (even) number of crossings of a reference line within half of the Brillouin zone identifies the nontrivial (trivial) phase.}
    \label{fig:wcc}
\end{figure}

\section{Wannier charge centers and the $\mathds{Z}_2$ invariant}
\label{sec:wann}

The maximally localized Wannier function associated with band index $n$ in the unit cell located at Bravais lattice vector $\mathbf{R}$ is defined as the Fourier transform of the Bloch states,
\begin{equation}
\ket{\mathbf{R}n}
=
\frac{V_{\mathrm{uc}}}{(2\pi)^d}
\int_{\mathrm{BZ}} d^d\mathbf{k}\;
e^{i\mathbf{k}\cdot\mathbf{R}}
\ket{\psi_{n\mathbf{k}}},
\label{eq:wann1}
\end{equation}
where $d$ is the spatial dimension, $V_{\mathrm{uc}}$ is the $d$-dimensional unit-cell volume (area in two dimensions), and $\ket{\psi_{n\mathbf{k}}}=e^{i\mathbf{k}\cdot\hat{\mathbf r}}\ket{u_{n\mathbf{k}}}$ with $\ket{u_{n\mathbf{k}}}$ the lattice-periodic Bloch function. The Wannier charge center (WCC) along $\hat x$ for the home cell $(\mathbf{R}=0)$ is
\begin{equation}
\bar x_n=\bra{\mathbf{0}n}\hat x\ket{\mathbf{0}n}.
\label{eq:wann2}
\end{equation}

In time-reversal-invariant systems, the occupied subspace can be organized into Kramers pairs labeled by $S=I,II$. For a given branch $S$, the Berry connection is
\begin{equation}
\mathbf{A}^S(\mathbf{k})
=
i\sum_{\alpha}
\bra{u^{S}_{\alpha\mathbf{k}}}\nabla_{\mathbf{k}}\ket{u^{S}_{\alpha\mathbf{k}}},
\label{eq:wann3}
\end{equation}
and the corresponding WCC can be written as the Brillouin-zone average
\begin{equation}
\bar x_n^S
=
\frac{V_{\mathrm{uc}}}{(2\pi)^d}
\int_{\mathrm{BZ}} d^d\mathbf{k}\;
\bra{u^{S}_{n\mathbf{k}}}
i\frac{\partial}{\partial k_x}
\ket{u^{S}_{n\mathbf{k}}}.
\label{eq:wann4}
\end{equation}

To diagnose the topological phase transition discussed in the main text, we track the WCC flow as a function of $k_y$ for two representative symmetry-lowered star-lattice realizations. Figure~\ref{fig:wcc}(a,b) shows the onsite-modulated model with $\delta_1=-0.08\,t$, $\delta_2=0.08\,t$, and $t_{\triangle}=t_{\bigtriangledown}=0.5\,t$ for two values of $\lambda_{\rm SO}$. As the SOC strength is increased from $0.04\,t$ to $0.06\,t$, the WCC flow changes from a topologically trivial to a nontrivial pattern, consistent with the band inversion at the $K$ point. Figure~\ref{fig:wcc}(c,d) shows the same analysis for the breathing star lattice with $t_{\triangle}=0.45\,t$ and $t_{\bigtriangledown}=0.5\,t$, where the transition occurs between $\lambda_{\rm SO}=0.03\,t$ and $0.04\,t$.

The $\mathds{Z}_2$ invariant is obtained by counting the number of WCC crossings with a reference line over half of the Brillouin zone~\cite{Vanderbilt,PhysRevB.84.075119}. An odd number of crossings signals a topologically nontrivial phase, whereas an even number indicates a trivial phase.
\begin{figure}
    \centering
    \includegraphics[width=1\linewidth]{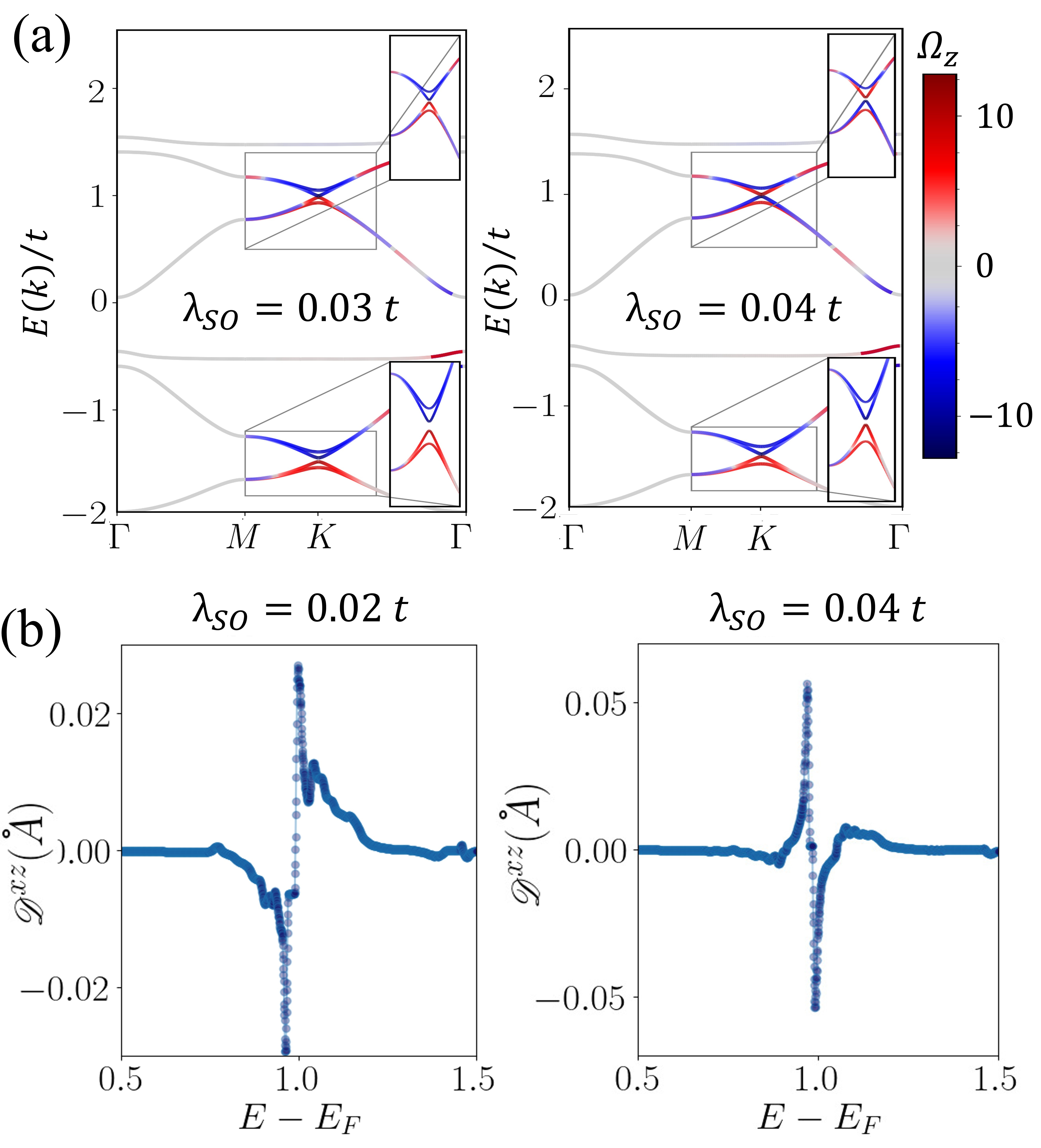}
    \caption{\textbf{Berry curvature and Berry-curvature dipole in the breathing star lattice.}
    (a) Berry-curvature-projected band structure for the breathing model with $t_{\triangle}=0.45\,t$ and $t_{\bigtriangledown}=0.5\,t$, shown for two SOC strengths.
    (b) Corresponding BCD in the presence of a small anisotropy $t_\perp\neq t$ representing $2\%$ tensile strain. The peak enhancement and sign reversal track the topological transition.}
    \label{fig:3}
\end{figure}

\section{Berry-curvature response in the breathing star lattice}
\label{sec:bc_bcd_supp}

In the main text, the onsite-modulated star lattice was used to illustrate the emergence of Berry-curvature hot spots and of a large BCD near the topological transition. The breathing star lattice shows the same qualitative physics and provides an independent route to inversion-symmetry breaking.

Figure~\ref{fig:3}(a) displays the Berry-curvature-projected band structure of the breathing model with $t_{\triangle}=0.45\,t$ and $t_{\bigtriangledown}=0.5\,t$ for two representative values of $\lambda_{\rm SO}$. Increasing $\lambda_{\rm SO}$ from $0.03\,t$ to $0.04\,t$ induces a band inversion near the $K$ point in the positive-energy sector. Correspondingly, the Berry-curvature hot spots at the avoided crossings reverse sign, reflecting the change in band character across the inversion.

The resulting BCD is shown in Fig.~\ref{fig:3}(b) for the same breathing model in the presence of a small inter-triangle anisotropy $t_\perp\neq t$, taken to represent a $2\%$ tensile strain. As in the onsite-modulated case, the dominant contribution is concentrated near the SOC-split Dirac crossings. The BCD therefore exhibits a pronounced enhancement close to the critical SOC strength and changes sign across the topological transition.

\begin{figure}
    \centering
    \includegraphics[width=0.9\linewidth]{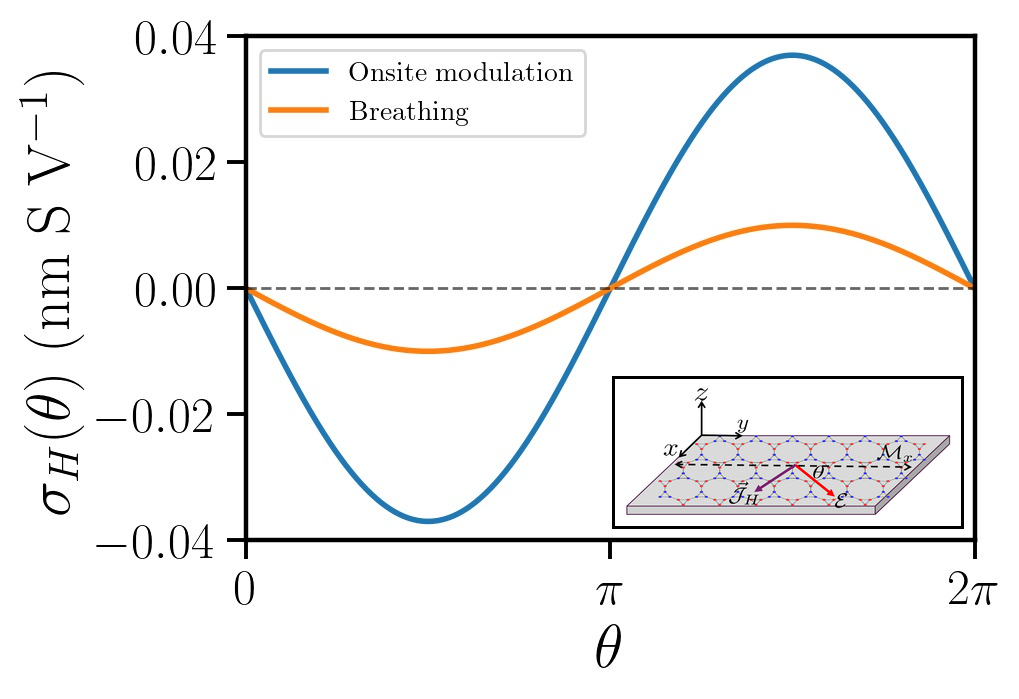}
    \caption{\textbf{Angular dependence of the nonlinear Hall conductivity.}
    Angular dependence of $\sigma_H(\theta)$ for the onsite-modulated and breathing realizations.
    Inset: geometry of the in-plane electric field $\boldsymbol{\mathcal E}$ (red arrow) applied at an angle $\theta$ relative to the mirror axis $\mathcal{M}_x$, generating a transverse nonlinear Hall current $\boldsymbol{\mathcal J}_H$ (blue arrow).}
    \label{fig:4}
\end{figure}
\section{Nonlinear Hall conductivity estimates}
\label{nlh}

To estimate the experimentally relevant NLHE response, we evaluate the angular dependence of the nonlinear Hall conductivity using Eq.~\eqref{eq:nlh_current} of the main text. Figure~\ref{fig:4} shows the result for two symmetry-lowered realizations of the star lattice: onsite modulation and breathing-type intra-triangle hopping modulation. In both cases, the response follows the mirror-constrained angular dependence expected for a system with a single remaining mirror plane $\mathcal{M}_x$.

For an in-plane electric field of magnitude $\mathcal{E}\sim10^4~\mathrm{V/m}$ applied at an angle $\theta$ relative to $\mathcal{M}_x$, the nonlinear Hall current takes the form $(\mathcal{J}_x,\mathcal{J}_y)=\mathcal{J}_H(\cos\theta,-\sin\theta)$~\cite{PhysRevB.107.205124}. Using the BCD values discussed in the main text and a representative relaxation time $\tau=10~\mathrm{ps}$, we obtain $\mathcal{J}_H\sim0.037~\mathrm{mA/cm}$ for the onsite-modulated case and $\mathcal{J}_H\sim0.01~\mathrm{mA/cm}$ for the breathing case. These values are within the reach of present experimental sensitivity.
\begin{figure*}
    \centering
    \includegraphics[width=0.9\linewidth]{Figures/supp_honeycomb.jpg}
    \caption{\textbf{Honeycomb-lattice description of the triphenyl--metal MOF.}
    (a) Structure of the triphenyl--Pb framework. Because the two Pb sites lie
    in different planes, a transverse electric field or substrate potential can
    break inversion symmetry.
    (b) DFT band structure of the pristine triphenyl--Pb framework, showing the
    characteristic honeycomb-lattice-like Dirac bands near the Fermi level. Here, for the pristine case $\tau_1$ = $\tau_2$ =$\tau_3$ = $\lambda/2$, where $\lambda$ denotes the effective nearest-neighbor hopping within the linker connecting the Pb atoms.}
    \label{fig:s5}
\end{figure*}

\section{Honeycomb-lattice downfolding of the triphenyl--Pb framework}
\label{sec:supp_honeycomb}

In addition to the star-lattice descriptions discussed in the main text, the
same downfolding philosophy can lead to a honeycomb-lattice effective model
when the orbital content near the Fermi level is distributed over the full
triphenyl--metal network. This situation is realized in the triphenyl--Pb
framework shown in Fig.~\ref{fig:s5}(a). In contrast to Cu--DCA, where the
metal orbitals contribute weakly near the Fermi level and can be integrated out
in favor of an effective star lattice, the low-energy states of the
triphenyl--Pb framework receive appreciable contributions from both the organic
linkers and the Pb atoms. The appropriate effective degrees of freedom are
therefore composite orbitals associated with the two sublattices of a
honeycomb network.

The pristine DFT band structure in Fig.~\ref{fig:s5}(b) displays the
characteristic Dirac dispersion of a honeycomb lattice near the Fermi level.
At the minimal level, this electronic structure can be described by a
nearest-neighbor honeycomb model,
\begin{equation}
H_{\rm hc}
=
\sum_{\langle ij\rangle}
t_{ij} c_i^\dagger c_j
+
i\lambda_{\rm SO}
\sum_{\langle\!\langle ij\rangle\!\rangle}
\nu_{ij} c_i^\dagger s_z c_j
+
\sum_i \Delta_i c_i^\dagger c_i ,
\end{equation}
where $t_{ij}$ denotes the effective nearest-neighbor hopping between the two
honeycomb sublattices, $\lambda_{\rm SO}$ is the intrinsic spin--orbit
coupling, and $\Delta_i$ describes inversion-symmetry-breaking onsite
potentials. The first term produces symmetry-protected Dirac cones, while SOC
and inversion-symmetry breaking gap these crossings and generate Berry-curvature
hot spots near the Fermi level.

Because the two Pb sites in the unit cell are vertically displaced, a
transverse electric field or substrate potential provides a natural route to
breaking inversion symmetry. Additional symmetry lowering, for example by
anisotropic linker functionalization, can reduce the remaining $C_3$ symmetry
and allow a finite Berry-curvature dipole. Thus, although the triphenyl--Pb
framework is more naturally described by a honeycomb rather than a star lattice,
it follows the same design principle for realizing nonlinear Hall responses in
MOFs.

\section{Methodology}
\label{sec:method}

\subsection{DFT methodology}
\label{sec:method_dft}

First-principles calculations were performed within density functional theory using the {\sc Quantum Espresso} package~\cite{giannozzi2009quantum,giannozzi2017advanced}. Exchange-correlation effects were treated within the Perdew--Burke--Ernzerhof generalized-gradient approximation~\cite{perdew1996generalized}, using the projector-augmented-wave framework~\cite{blochl1994projector}. Scalar-relativistic and fully relativistic pseudopotentials were employed for calculations without and with SOC, respectively. We used a kinetic-energy cutoff of 46~Ry and an $8\times8\times1$ Monkhorst--Pack $k$ mesh~\cite{monkhorst1976special}. A vacuum spacing of $10~\ang$ was introduced along the out-of-plane direction to suppress interlayer interactions. Topological properties and Berry-curvature-related observables were additionally benchmarked using the \texttt{WannierBerri} package~\cite{Tsirkin2021WannierBerri}, interfaced with \texttt{PythTB}~\cite{Coh_Vanderbilt_2022_PythTB}.

\subsection{Real-space decimation}
\label{sec:method_decimation}

The tight-binding form of the Schr\"odinger equation is
\begin{equation}
(E-\varepsilon_i)\phi_i=\sum_j t_{ij}\phi_j,
\label{eq:difference1}
\end{equation}
where $E$ is the eigenenergy, $\varepsilon_i$ is the onsite potential on site $i$, $\phi_i$ is the corresponding wave-function amplitude, and $t_{ij}$ is the hopping amplitude between sites $i$ and $j$.

For the linker segment connecting neighboring effective triangles [Fig.~\ref{fig:1}(c) of the main text], the local difference equations for the site pair $(i,j)$ can be written as
\begin{equation}
(E-\varepsilon)\phi_{i/j}=\lambda\,\phi_{j/i}+\lambda\,\phi_{g/h},
\label{eq:diff1}
\end{equation}
where $\lambda$ denotes the nearest-neighbor hopping within the linker. Substituting Eq.~\eqref{eq:diff1} into the corresponding equation for the next pair $(g,h)$,
\begin{equation}
(E-\varepsilon)\phi_{g/h}=\lambda\,\phi_{i/j}+\lambda\,\phi_{e/f},
\label{eq:diff2}
\end{equation}
and taking the low-energy condition $E-\varepsilon\approx 0$, one obtains
\begin{equation}
(E-\varepsilon)\phi_{g/h}=-\lambda\,\phi_{h/g}+\lambda\,\phi_{e/f}.
\label{eq:diff3}
\end{equation}
This identifies the first renormalized hopping as $u=-\lambda$.

Iterating the same decimation step through the sequence of pairs $\{(g,h),(e,f),(c,d),(a,b)\}$ reduces the full linker segment to an effective bond between the terminal sites $L$ and $R$ shown in Fig.~\ref{fig:1}(c) of the main text. The resulting renormalized parameters are
\begin{equation}
v=2\lambda,\qquad
w=-\lambda,\qquad
x=\lambda,\qquad
t=-\frac{{\lambda'}^2}{\lambda}.
\label{eq:renormalized}
\end{equation}
This analytical decimation establishes the effective star-lattice description used throughout the main text and provides the basis for relating the chemistry of the MOF linkers to the symmetry-selective hopping amplitudes of the low-energy model.










\end{document}